\documentclass[prl,twocolumn,showpacs,amsmath,superscriptaddress,psfig,amssymb]{revtex4}
\usepackage{graphicx}% Include figure files
\usepackage{dcolumn}% Align table columns on decimal point
\usepackage{bm}% bold math

\begin{document}

%\preprint{APS/123-QED}

\title{Superconductivity at 32K and anisotropy in Tl$_{0.58}$Rb$_{0.42}$Fe$_{1.72}$Se$_{2}$ crystals }

\author{Hangdong Wang}
\affiliation {Department of Physics, Zhejiang University, Hangzhou 310027, China}
\affiliation {Department of Physics, Hangzhou Normal University, Hangzhou 310036, China}

\author{Chihen Dong}
\affiliation {Department of Physics, Zhejiang University, Hangzhou 310027, China}

\author{Zujuan Li}
\affiliation {Department of Physics, Zhejiang University, Hangzhou 310027, China}

\author{Shasha Zhu}
\affiliation {Department of Physics, Zhejiang University, Hangzhou 310027, China}

\author{Qianhui Mao}
\affiliation {Department of Physics, Zhejiang University, Hangzhou 310027, China}

\author{Chunmu Feng}
\affiliation {Department of Physics, Zhejiang University, Hangzhou 310027, China}

\author{H. Q. Yuan}
\affiliation {Department of Physics, Zhejiang University, Hangzhou 310027, China}

\author{Minghu Fang}
\email{mhfang@zju.edu.cn}
\affiliation {Department of Physics, Zhejiang University, Hangzhou 310027, China}

\date{\today}% It is always \today, today,

\begin{abstract}
\noindent Single crystals of
Tl$_{0.58}$Rb$_{0.42}$Fe$_{1.72}$Se$_{2}$ are successfully grown
with the superconducting transition temperatures
\textit{T$_{c}$$^{onset}$}=32K and \textit{T$_c$$^{zero}$}=31.4K.
The Hall coefficient exhibits a multi-band behavior, which is very
similar to that of all other Fe-based superconductors. We found that
the susceptibility at the normal state decreases with decreasing the
temperature, indicating a strong antiferromagnetic (AFM) spin
fluctuation at the normal state, which might be related to the
superconductivity (SC). We also determined the upper critical fields
in \textit{ab}-plane and along \textit{c}-axis. The anisotropy of
the superconductivity determined by the ratio of
\textit{H$_{c2}$$^{ab}$} and \textit{H$_{c2}$$^c$} is estimated to
5.0, which is larger than that in (Ba,K)Fe$_{2}$As$_2$ and
BaFe$_{2-x}$Co$_x$As$_2$, but smaller than that in cuprate superconductors.

\end{abstract}

\maketitle

Since superconductivity at 26K in LaO$_{1-x}$F$_x$FeAs was
discovered by Kamihara \textit{et al.}\cite{Kamihara 2008}, the
iron-based superconductors have received worldwide attention in the
last three years. There have been four major types of iron-based
compounds reported to exhibit SC after doping or under high
pressures, i.e. 1111-type ReFeAsO (Re = rare earth)\cite{Kamihara
2008,Chen 2008,GFChen 2008,Ren 2008,Wen 2008,Wang C 2008}, 122-type
BFe$_2$As$_2$ (B=Ba, Sr or Ca)\cite{Rotter 2008,Sefat 2008,Li LJ
2009}, 111-type AFeAs (A = alkali metal)\cite{Wang XC 2008}, and
11-type tetragonal FeTe$_{1-x}$Se(S)$_x$\cite{Hsu 2008,Fang 2008}.
All these iron-based superconductors share a common layered
structure based on a square planar Fe$^{2+}$ layer, tetrahedrally
coordinated pnictogen (P, As) or chalcogen (S, Se, Te) anions. The
emergence of superconductivity upon destruction of long-range
antiferromagnetic (AFM) order is qualitatively similar to
observations in the layered cuprate high temperature
superconductors. However, in the cuprates, the repulsive interaction
among the electrons is so strong that the parent compounds are Mott
insulators. By contrast, all iron-based parents above mentioned are
metallic.

Very recently, superconductivity at about 30K was reported in
K$_x$Fe$_2$Se$_2$ \cite{Guo 2010} and Cs$_{0.8}$Fe$_2$Se$_{1.96}$
\cite{Krzton 2010}. Our group \cite{Fang 2010} have confirmed that
superconductivity at 31K occurring in the newly (Tl,K)Fe$_x$Se$_2$
compound (denoted as 122-type iron-chalcogenide) is also in
proximity of an AFM insulator. We found that (Tl,K)Fe$_{1.5}$Se$_2$
with a Fe-vacancy super-lattice is an AFM insulator with the Neel
temperature, \textit{T}$_N$=250 K, which can be regarded as the
parent of this system. With increasing the Fe-content, the AFM order
is reduced. When the magnetism is eliminated, superconductivity at
31K emerges. In this letter, we report the successful growing of the
single crystals for another new compound
Tl$_{0.58}$Rb$_{0.42}$Fe$_{1.72}$Se$_2$. The onset and
zero-resistivity transition temperature were estimated to be 32 K
and 31.4 K, respectively. We found that the Hall coefficient
exhibits a multi-band behavior. The susceptibility result indicates
a strong AFM spin fluctuation at the normal state. We also
determined the upper critical fields in \textit{ab}-plane and
along\textit{c}-axis. The anisotropy of the superconductivity
determined by the ratio of \textit{H}$_{c2}$$^{ab}$ and
\textit{H}$_{c2}$$^c$ is estimated to be 5.0.

Single crystals were grown by the Bridgeman method. First, Rb$_2$Se,
Tl$_2$Se, Fe and Se powders with high purity (99.99\%) were mixed in
an appropriate stoichiometry and were put into alumina crucibles and
sealed in evacuated silica tube. The mixture was heated up to
950$^o$C and held for 6 hours. Then the melting mixture was cooled
down to 700$^o$C in the cooling rate of 3$^o$C/h and finally the
furnace was cooled to room temperature with the power shut off. The
obtained single crystals show the flat shiny surface and are easy to
cleave. The composition of crystals,
Tl$_{0.58}$Rb$_{0.42}$Fe$_{1.72}$Se$_2$, was determined by using an
Energy Dispersive X-ray Spectrometer (EDXS). The structure of single
crystals was characterized by X-ray diffraction (XRD). Magnetic
susceptibility measurements were carried out using the
\textit{Quantum Design} MPMS-SQUID. The measurements of resistivity,
Hall effect and magneto-resistance were done using the
\textit{Quantum Design} Physical Properties Measurement System
PPMS-9.

Figure 1 shows the XRD (Fig.1a) pattern of powder obtained by
grounding the crystals and single crystal XRD (Fig.1b) pattern for
Tl$_{0.58}$Rb$_{0.42}$Fe$_{1.72}$Se$_2$. Most peaks in the powder
XRD pattern can be well indexed with a ThCr$_2$Si$_2$-type structure
(space group: \textit{I}4/\textit{mmm}). The lattice parameters,
\textit{a} = 3.896{\AA}, and \textit{c} = 14.303{\AA} was obtained
by the fitting XRD data. One small peak marked by a triangle belongs
to FeSe compound due to small amount existing between crystals. Only
(00\textit{l}) peaks were observed in the single crystal XRD
pattern, indicating that the crystallographic \textit{c} axis is
perpendicular to the plane of the single crystal. The interesting is
that there is another series of (00\textit{l}) peaks (marked by the
asterisks) with \textit{c} = 14.806{\AA}, which is larger than that
of the main phase, implying that there may be a modulation structure
along \textit{c} axis due to the existence of Fe-vacancy. But there
is only a small peak appearing in the powder XRD pattern. This
modulation structure was also observed in Cs$_x$Fe$_2$Se$_2$
crystals\cite{Krzton 2010}.

\begin{figure}
  \includegraphics[width=8cm]{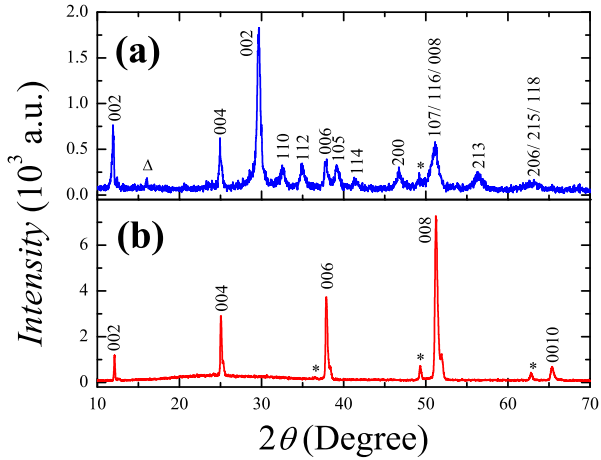}\\
    \caption{(Color online)(a) Powder and (b) Single crystal X-ray diffraction pattern of Tl$_{0.58}$Rb$_{0.42}$Fe$_{1.72}$Se$_2$}\label{}
\end{figure}

Figure 2(a) shows the resistivity in \textit{ab}-plane,
$\rho$$_{ab}$(T), and along \textit{c} axis, $\rho$$_c$(T), as a
function of temperature for Tl$_{0.58}$Rb$_{0.42}$Fe$_{1.72}$Se$_2$
crystal. At higher temperatures, both $\rho$$_{ab}$(T) and
$\rho$$_c$(T) exhibit a semiconductor-like behavior, and displays a
maximum resistivity at about 150 K, and shows a metallic behavior
below 150 K. A metal-insulator transition occurs at about 150K.
Similar behavior of $\rho$$_{ab}$(T) was also observed in
K$_x$Fe$_2$Se$_2$ \cite{Guo 2010}, Cs$_x$Fe$_2$Se$_2$ \cite{Krzton
2010}, (Tl,K)Fe$_x$Se$_2$\cite{Fang 2010} and Rb$_x$Fe$_2$Se$_2$
\cite{Ying 2010,Wang 2010} crystals. We identified that the
temperature of the maximum resistivity in $\rho$$_{ab}$(T) depends
on the actual Fe-content in the crystals for the similar system
(Tl,K)Fe$_x$Se$_2$ \cite{Fang 2010}. It should be noted that the
absolute value of resistivity at the normal state is quite large,
for example, $\rho$$_{ab}$ is of 10 m$\Omega$ cm at 300 K, which is
much larger than that in other typical iron-based superconductors.
This may be attributed to the this compound residing in the
under-doping region in the phase diagram, as discussed in
(Tl,K)Fe$_x$Se$_2$ system by us.\cite{Fang 2010} Another, from the
ratio value of $\rho$$_c$/$\rho$$_{ab}$=30-45 estimated from the
resistivity data at the normal state, the anisotropy is smaller than
that in (Tl,K)Fe$_x$Se$_2$ system, where
$\rho$$_c$/$\rho$$_{ab}$=70-80 \cite{Fang 2010}.

\begin{figure}
  \includegraphics[width=8cm]{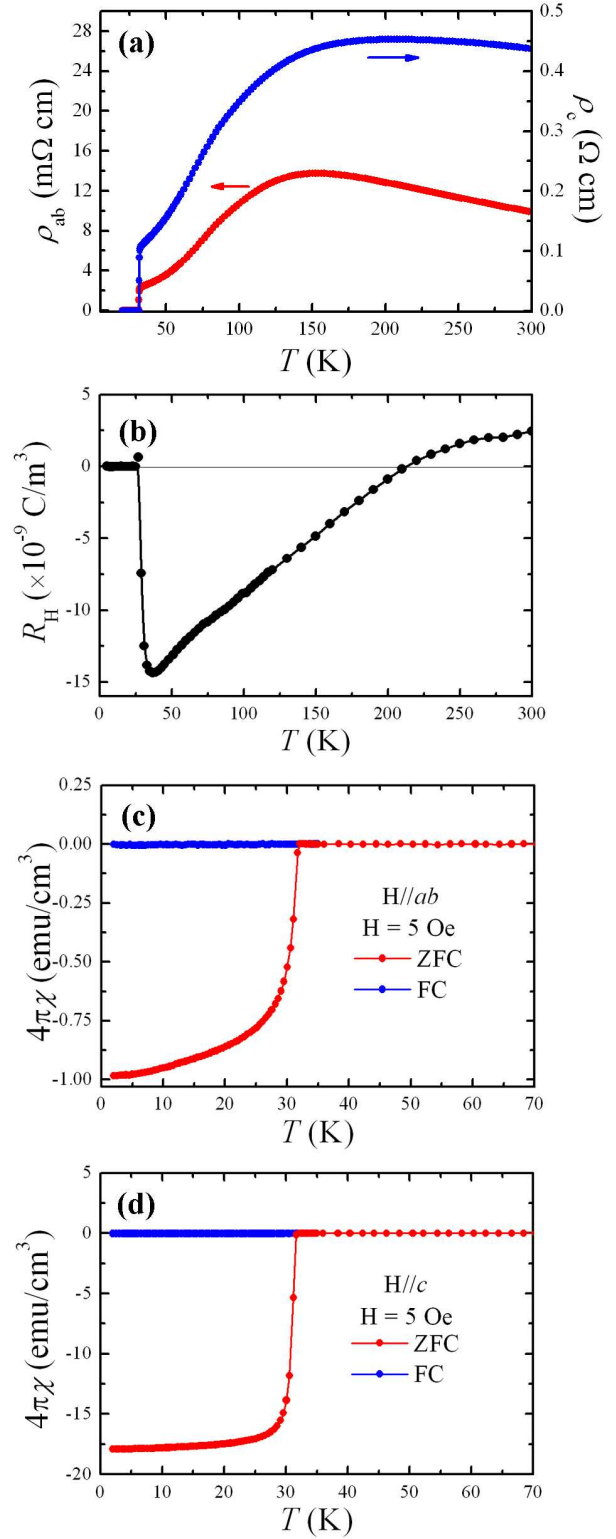}\\
  \caption{(Color online)(a) Temperature dependence of \textit{ab}-plane and \textit{c}-axis resistivity. (b)Temperature dependence of Hall coefficient. (c) Temperature dependence of dc magnetic susceptibility for both zero field cooling (ZFC) and field cooling processes (FC) at a magnetic field of \textit{H}= 5 Oe applied along c axis and (d) in \textit{ab}-plane for Tl$_{0.58}$Rb$_{0.42}$Fe$_{1.72}$Se$_2$ crystal.}\label{}
\end{figure}

In order to explore the electronic structure in this compound, we
measured the Hall coefficient \textit{R$_H$}. Figure 2(b) shows the
temperature dependence of Hall coefficient, \textit{R}$_H$(T),
forTl$_{0.58}$Rb$_{0.42}$Fe$_{1.72}$Se$_{2}$ crystal. The
interesting is that the change in the signal of \textit{R}$_H$(T) at
214 K occurs from positive to negative, which is the typical
behavior of the multi-band of the electronic band and a common
feature of Fe-based superconductors. Above 214K, the positive value
of \textit{R}$_H$ implies that the carrier in this compound is
dominated by holes. The negative value of \textit{R}$_H$ from 214K
to \textit{T}$_c$ indicates that the carrier in the compound is
dominated by electrons. This result means that there are both hole
pocket and electron pocket near the Fermi energy, which is very
similar to that of all other Fe-based superconductors although there
are Fe vacancies on the Fe-planar in
Tl$_{0.58}$Rb$_{0.42}$Fe$_{1.72}$Se$_{2}$ compound. In the
(Tl,K)Fe$_x$Se$_2$ system \cite{Fang 2010}, we found that no signal
change of \textit{R$_H$} occurs, indicating that the carrier is
dominated by electrons from \textit{T$_c$} to 350 K. The difference
of the electron structure between both may be due to the difference
between K and Rb radii.

From both  $\rho$$_{ab}$(T) and $\rho$$_{c}$(T), we can see that a
sharp superconducting transition occurs at
\textit{T$_c$$^{onset}$}=32 K, \textit{T$_c$$^{mid}$}=31.5 K and
\textit{T$_c$$^{zero}$}=31.4 K , confirmed by the existence of
diamagnetism below \textit{T$_c$} (see Fig. 2c and 2d). As the
magnetic field is applied in \textit{c} axis, the diamagnetic signal
at the superconducting state is very large and over negative unity
due to larger demagnetization factor. When the magnetic field is
applied in \textit{ab} plane, the demagnetization effect can being
ignored, diamagnetic signal reaches to 98\% at low temperatures. All
of these demonstrate a bulk SC emerging in the
Tl$_{0.58}$Rb$_{0.42}$Fe$_{1.72}$Se$_2$ crystal, which is different
than that observed in (Tl,K)Fe$_x$Se$_2$ by us \cite{Fang 2010},
where no bulk SC was observed in the range of 1.70 $\leq$ $x$ $<$
1.78.

\begin{figure}
  \includegraphics[width=8cm]{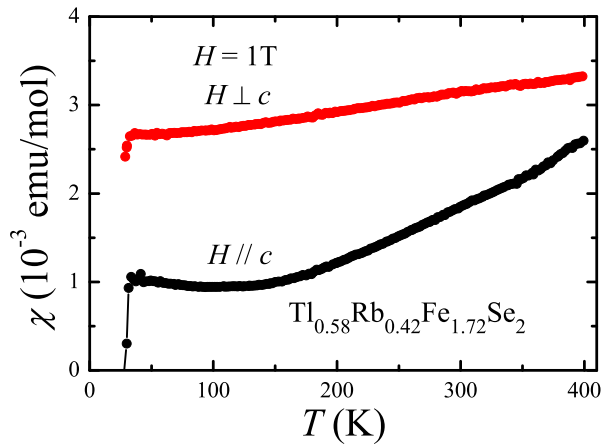}\\
    \caption{(Color online) The magnetic susceptibility measured at 1 Tesla for Tl$_{0.58}$Rb$_{0.42}$Fe$_{1.72}$Se$_2$ crystal with the magnetic field along and perpendicular to \textit{c}-axis.}\label{}
\end{figure}

Figure 3 shows the magnetic susceptibility at the normal state as a
function of temperature, $\chi$$_{c}$(T) and $\chi$$_{ab}$(T), for
the Tl$_{0.58}$Rb$_{0.42}$Fe$_{1.72}$Se$_2$ crystal with a magnetic
field of 1 T applied parallel and perpendicular to the
\textit{c}-axis. A drop in susceptibility at about 30 K corresponds
to the superconducting transition. When the magnetic field was
applied along \textit{c}-axis, the susceptibility,  $\chi$$_{c}$(T)
value above 150 K decreases linearly with decreasing the
temperature, then shows a little increase until the superconducting
transition occurs. When the magnetic field was applied in
\textit{ab}-plane, the  $\chi$$_{ab}$ value decreases well linearly
with decreasing the temperature in the temperature range of 30K
$\leq$ T $\leq$ 400K. The behavior of both  $\chi$$_{ab}$ and
$\chi$$_{c}$ drop with decreasing the temperature implies a strong
AFM spin fluctuation at the normal state, which might be related to
the SC.

\begin{figure}
  \includegraphics[width=8cm]{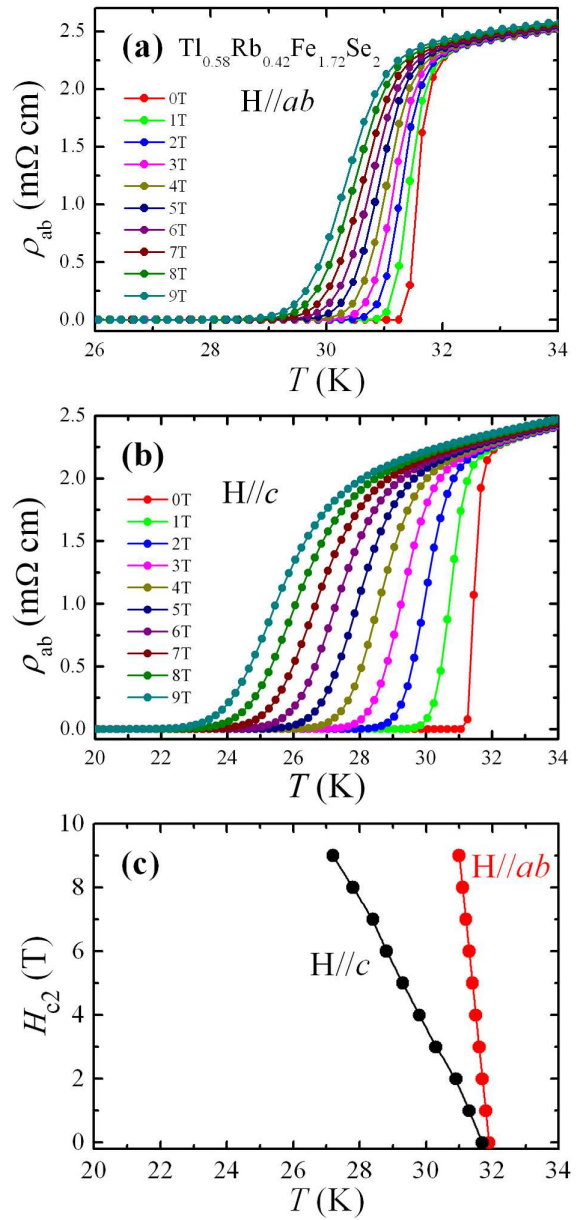}\\
    \caption{(Color online) The temperature dependence of the resistivity with magnetic field (a) parallel and (b) perpendicular to the \textit{ab}-plane, respectively. (c) The temperature dependence of upper critical field \textit{H$_{c2}$}(T) for Tl$_{0.58}$Rb$_{0.42}$Fe$_{1.72}$Se$_2$ crystal}\label{}
\end{figure}

The temperature dependence of resistivity from 20K to 34K with
different magnetic fields applied along \textit{c}-axis and in
\textit{ab}-plane is shown in Fig. 4(a) and 4(b). We defined the
\textit{T}$_c$ as the temperature where the resistivity was 90\%
drop at the superconducting transition to determine the upper
critical fields, \textit{H$_{c2}$}. The \textit{H$_{c2}$(T)}
determined in this way is shown in Fig. 4(c). The
\textit{H$_{c2}$(T)} exhibits a rather linear temperature dependence
for both directions. Thus we can easily get the value of the slope
for both directions: \textit{-dH$_{c2}$$^{ab}$/dT$\mid$T$_c$ }=10.0
T/K , \textit{-dH$_{c2}$$^{c}$/dT$\mid$T$_c$ }=2.0 T/K. The
\textit{H$_{c2}$(0)} value can be estimated by the
Werthamer-Helfand-Hohenberg (WHH) equation \cite{Werth 1966}
\textit{H$_{c2}$(T)}=0.693[-(\textit{dH$_{c2}$/dT})]\textit{T$_c$}
with \textit{T$_c$}=32 K, to be 221 T and 44.2 T with the magnetic
field applied in \textit{ab}-plane and along \textit{c}-axis,
respectively. The anisotropy \textit{H$_{c2}$$^{ab}$(0)/
H$_{c2}$$^c$(0)} is about 5.0, which is larger than that in
Ba$_{0.6}$K$_{0.4}$Fe$_2$As$_2$\cite{Wen HH 2008} and
Ba(Fe$_{0.92}$Co$_{0.08}$)$_2$As$_2$ \cite{Ni 2008-2}.

In summary, we successfully grew the single crystals of
Tl$_{0.58}$Rb$_{0.42}$Fe$_{1.72}$Se$_2$ with the superconducting
transition temperature \textit{T$_c$$^{onset}$}=32 K and
\textit{T$_c$$^{zero}$}=31.4 K. The Hall coefficient exhibits a
multi-band behavior, which is very similar to that of all other
Fe-based superconductors although there are Fe vacancies on the
Fe-planar in Tl$_{0.58}$Rb$_{0.42}$Fe$_{1.72}$Se$_2$ compound. We
also determined the upper critical fields in \textit{ab}-plane and
along \textit{c}-axis. The anisotropy of the superconductivity
determined by the ratio of \textit{H$_{c2}$$^{ab}$} and
\textit{H$_{c2}$$^c$} is estimated to 5.0, which is larger than that
in Ba$_{0.6}$K$_{0.4}$Fe$_2$As$_2$ and
BaFe$_{1.92}$Co$_{0.08}$As$_2$, but smaller than that in cuprate
superconductors.

This work is supported by the Nature Science Foundation of China
(Grant No. 10974175 and 10934005), the National Basic Research
Program of China (973 Program) under grant No. 2011CBA00103 and
2009CB929104, and PCSIRT of the Ministry of Education of China
(Grant No. IRT0754).

\end{document}